\documentclass[12pt]{iopart}
\usepackage{iopams,epsf,psfig}

\begin{document}
\title[Quantised Three-Pillar Problem]{Quantised Three-Pillar Problem}

\author[Bender, Brody and Meister]{
Carl M. Bender$^{*}$, Dorje C. Brody$^{\dagger}$ and Bernhard K.
Meister$^{\dagger}$}

\address{$*$Department of Physics, Washington University, St.
Louis MO 63130, USA}
\address{$\dagger$Blackett Laboratory, Imperial College, London
SW7 2BZ, UK }

\begin{abstract}
This paper examines the quantum mechanical system that arises when
one quantises a classical mechanical configuration described by an
underdetermined system of equations. Specifically, we consider the
well-known problem in classical mechanics in which a beam is
supported by three identical rigid pillars. For this problem it is
not possible to calculate uniquely the forces supplied by each
pillar. However, if the pillars are replaced by springs, then the
forces are uniquely determined. The three-pillar problem and its
associated indeterminacy is recovered in the limit as the spring
constant tends to infinity. In this paper the spring version of
the problem is quantised as a constrained dynamical system. It is
then shown that as the spring constant becomes large, the quantum
analog of the ambiguity reemerges as a kind of quantum anomaly.
\end{abstract}

\submitto{\JPA}

\vskip1pc

\begin{figure}[t]
\begin{center}
{\centerline{\psfig{file=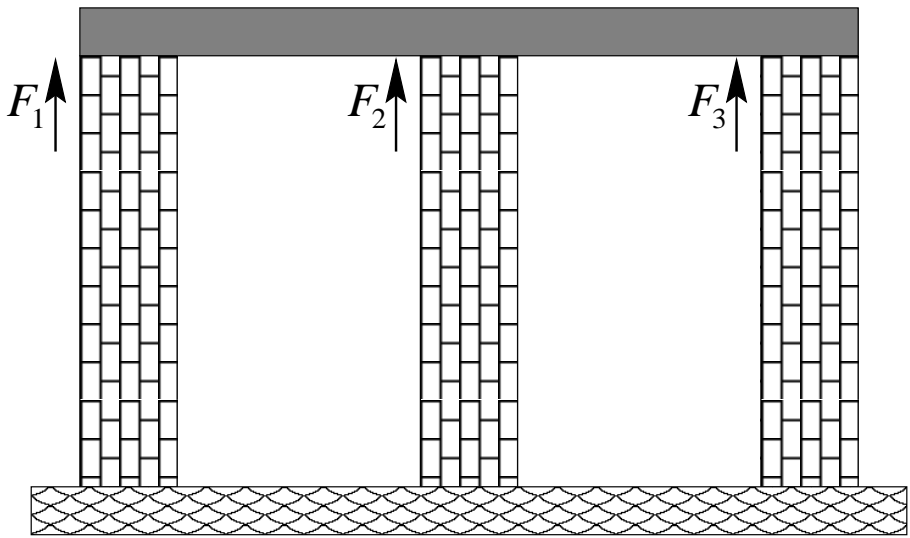,width=12cm,angle=270}}}
\end{center}
\caption{ \label{F1} Three-pillar problem: an elementary example
of a classical mechanical system characterised by an
underdetermined set of equations. A beam in static equilibrium is
supported by three identical incompressible pillars. The condition
of static equilibrium is described by {\sl two} equations, the
force balance and the torque balance. However, there are {\sl
three} unknowns, the upward forces $F_1$, $F_2$, and $F_3$
provided by the three pillars. As a consequence, the forces cannot
be determined by the equations of classical mechanics.}
\end{figure}

Consider a rigid beam of weight $W$ supported by three identical
incompressible pillars (see Fig.~\ref{F1}). Let the upward forces
provided by the pillars be $F_1$, $F_2$, and $F_3$, respectively.
If the beam is in static equilibrium, then the sum of the upward
forces must equal $W$:
\begin{eqnarray}
F_1+F_2+F_3=W. \label{eq1}
\end{eqnarray}
Also, the torque on the beam about any point must vanish. If we
calculate the torque about the centre, we obtain the condition
\begin{eqnarray}
F_1=F_3. \label{eq2}
\end{eqnarray}
(Calculating the torque about any other point does not give
additional information). The simultaneous solution to (\ref{eq1})
and (\ref{eq2}) is
\begin{eqnarray}
F_1&=&{1\over2}(W-f),\nonumber\\
F_2&=&f,\nonumber\\
F_3&=&{1\over2}(W-f), \label{eq3}
\end{eqnarray}
where $f$ is an arbitrary force that cannot be determined by the
conditions of the problem. This is an elementary example of a
classical mechanical system whose physical characteristics are
underdetermined.

\begin{figure}[t]
\begin{center}
{\centerline{\psfig{file=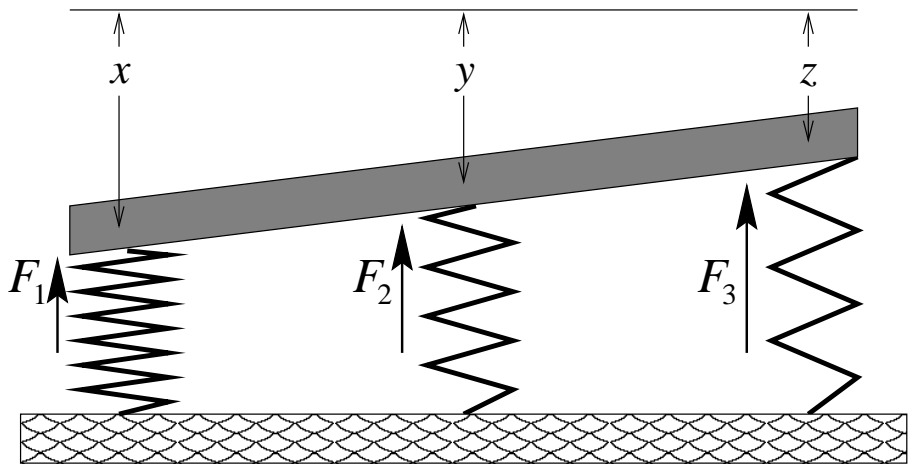,width=12cm,angle=270}}}
\end{center}
\caption{ \label{F2} Three spring problem: the rigid pillars in
Fig.~\ref{F1} are replaced by identical springs having spring
constant $k$. The classical equations describing the system in
static equilibrium have a unique solution. The beam is shown
resting on the springs whose displacements from equilibrium are
$x$, $y$, and $z$. The displacements are determined to be
$x=y=z=W/(3k)$. Thus, the forces provided by the springs are
uniquely determined to be $W/3$ and the ambiguity in the
three-pillar problem is removed.}
\end{figure}

It is possible to reformulate the three-pillar problem in such a
way as to remove this ambiguity. We replace the three pillars by
three identical springs having spring constant $k$. Now, the rigid
beam rests on these springs, and the springs are displaced from
their equilibrium lengths by the amounts $x$, $y$, and $z$ (see
Fig.~\ref{F2}). For this problem we can determine $x$, $y$, and
$z$, and thus we can determine the forces uniquely. The force
balance equation reads
\begin{eqnarray}
kx+ky+kz=W, \label{eq4}
\end{eqnarray}
and the torque balance condition implies that
\begin{eqnarray}
x=z. \label{eq5}
\end{eqnarray}
The condition that the beam is straight and rigid implies that
\begin{eqnarray}
x+z=2y, \label{eq6}
\end{eqnarray}
which is a new equation having no analogue in the three-pillar
problem. The simultaneous solution of equations (\ref{eq4}) --
(\ref{eq6}) gives $x=y=z={W\over3k}$, and thus the upward forces
imposed on the beam by the three springs are
\begin{eqnarray}
F_1={1\over3}W,\qquad F_2={1\over3}W,\qquad F_3={1\over3}W.
\label{eq7}
\end{eqnarray}

The ambiguity in the three-pillar problem is eliminated because
the flexibility in the springs and the rigidity of the beam give
rise to the additional condition (\ref{eq6}) that allows us to
solve for the forces. Thus, if the pillars are replaced by
compressible objects, which need not even be identical, then the
indeterminacy is lifted. However, if we take the limit
$k\to\infty$ in the spring problem, then the springs become rigid
objects, and we recover the three-pillar problem. To see this, we
note from (\ref{eq4}) that the limit $k\to\infty$ gives
\begin{eqnarray}
x+y+z=0 \label{eq8}
\end{eqnarray}
because $W$ is a constant. The torque condition (\ref{eq5}) is
still valid, and along with (\ref{eq6}) we find that $x=y=z=0$. We
obtain this result because in this limit the springs become
incompressible, and the deviation from the equilibrium length of
the springs must vanish. However, the forces are no longer
determined because they are expressed in the ambiguous form
$F_i=\infty\times0$ ($i=1,2,3$). Thus, any values for the forces
are allowed subject to constraints (\ref{eq1}) and (\ref{eq2}).

Let us now consider the quantum mechanical version of the
three-spring problem. To describe the quantised three-spring
problem we start with the Hamiltonian
\begin{eqnarray}
H={1\over2m}\left(p_x^2+p_y^2+p_z^2\right)+{1\over2}k\left(x^2+
y^2+z^2\right),
\label{eq9}
\end{eqnarray}
which represents three uncoupled harmonic oscillators. Note that
we have shifted the zeros of the variables $x$, $y$, and $z$ by
the amount ${W\over3k}$ so that the beam oscillates about its
classical resting position. We then impose the constraint
(\ref{eq6}) to eliminate the dynamical variable $y$. The
Hamiltonian thus obtained is
\begin{eqnarray}
H={1\over2m}\left[p_x^2+{1\over2}\left(p_x+p_z\right)^2+p_z^2\right]+
{1\over2}k\left[x^2+{1\over4}(x+z)^2+z^2\right]. \label{eq10}
\end{eqnarray}

It is convenient to make the change of variables $r=x+z$ and
$s=x-z$. The variable $r$ represents vertical oscillatory motion
of the centre of mass of the beam and the variable $s$ is
associated with the rotational motion of the beam. In terms of
these variables, the resulting Hamiltonian is diagonal and we
obtain the time-independent Schr\"odinger equation
\begin{eqnarray}
\left[-{1\over2m}\left( 3{\partial^2\over\partial r^2}
+2{\partial^2\over\partial s^2} \right)+{1\over8}k\left(3r^2+2s^2
\right) \right] \psi(r,s)=E\psi(r,s). \label{eq11}
\end{eqnarray}

Let us calculate the expectation value of the force, given by the
negative derivative of the potential, applied by the central
oscillator. We perform this calculation in the ground state of the
quantum system. The (unnormalised) ground-state wave function is
\begin{eqnarray}
\psi_0(r,s)=\exp\left(-{1\over4}\sqrt{km}\,r^2\right) \exp
\left(-{1\over4}\sqrt{km}\,s^2\right). \label{eq12}
\end{eqnarray}
The expectation value of force in the central oscillator, taking
into account the shift of the zero, is given by $W/3$ plus the
integral
\begin{eqnarray}
\langle0|{\rm Force}|0\rangle &=& -\frac{k}{2}{\int\int {\rm d}
r\,r\,{\rm d}s \,\exp\left[-{1\over2}\sqrt{mk}\left(r^2+s^2\right)
\right] \over \int\int {\rm d}r\,{\rm d}s\,\exp
\left[-{1\over2}\sqrt{mk} \left(r^2+s^2\right)\right]} \nonumber
\\ &=& -\frac{k}{2}{\int {\rm d}r\,r\, \exp
\left(-{1\over2}\sqrt{mk}\,r^2\right) \over \int {\rm d} r\,
\exp\left(-{1\over2}\sqrt{mk}\,r^2\right)}, \label{eq13}
\end{eqnarray}
where we have divided out the integral over the variable $s$. So
long as the spring constant $k$ is finite, a reflection symmetry
argument implies that the integral in the numerator vanishes.
Indeed, in {\sl any} state, the expectation value of $r$ vanishes
by reflection symmetry. However, as $k\to\infty$, the expression
in (\ref{eq13}) is the product of 0 and $\infty$, and we have
obtained an ambiguous result.

This ambiguity may be regarded as a kind of quantum anomaly.
Typically, in quantum field theory one encounters anomalies for
which there is a divergent integral multiplying a quantity that
vanishes because of a geometrical symmetry argument. For example,
in the Schwinger model of two-dimensional electrodynamics the
trace of the photon propagator formally vanishes by the symmetry
properties of two-dimensional gamma matrices \cite{schwinger}.
However, the photon propagator contains an integral that is
logarithmically divergent. There are many tricks that can be used
to {\sl regulate} and then calculate such ambiguous quantities. In
the case of the Schwinger model one can regulate the divergent
integral by performing the calculation in $2+\epsilon$ dimensions;
in $2+\epsilon$ dimensions the trace of the gamma matrices does
not vanish and the integral is finite. One then takes the limit
$\epsilon\to0$ to calculate the value of the anomaly. Another
example of an anomaly is the so-called trace anomaly, where the
stress-energy tensor formally has a vanishing trace, but is
represented by a divergent integral \cite{coleman}.

We view the expression in (\ref{eq13}) as a kind of quantum
anomaly because there is one quantity (here, the integral) that
vanishes by a geometrical symmetry argument (reflection of $r$).
This quantity is multiplied by a second factor $k$ that diverges
(here, $k\to\infty$). Our objective is to demonstrate that,
depending on the regularisation scheme chosen, we can get any
result for the expectation value (\ref{eq13}) of the force.

As an example, we can regulate the anomalous ambiguous product
\begin{eqnarray}
A=\lim_{\Lambda\to\infty}\Lambda\int_{-\infty}^\infty {\rm d}r\,r
e^{-r^2} \label{eee}
\end{eqnarray}
as follows:
\begin{eqnarray}
A\equiv\lim_{\Lambda\to\infty}\Lambda\int_{-\infty}^{\sqrt{\log
[\Lambda/(2a)]}} {\rm d}r\,r e^{-r^2}, \label{eq14}
\end{eqnarray}
where $a>0$ is arbitrary. Evaluating (\ref{eq14}) exactly, we
obtain
\begin{eqnarray}
A= -a. \label{eq15}
\end{eqnarray}
Apparently, by choosing an appropriate regulation scheme we can
obtain any value $-a$ for $A$. Since the expression for the force
in (\ref{eq13}) has the form of the anomalous product in
(\ref{eee}), we see that as the spring constant $k$ tends to
infinity, the expectation value of the force is arbitrary.

\vspace{0.5cm}
\begin{footnotesize}
We wish to express our gratitude to H.~F.~Jones for his
stimulating suggestions. DCB gratefully acknowledges financial
support from The Royal Society. This work was supported in part by
the U.S.~Department of Energy.
\end{footnotesize}

\end{document}